\documentclass[letterpaper,12pt]{article}

\usepackage{mathptmx}
\usepackage{amsmath}
\usepackage[authoryear,comma,longnamesfirst,sectionbib]{natbib} 
\usepackage{graphics}
\usepackage{graphicx}

\begin{document}


\vspace*{0.5cm}
\begin{center}
\textbf{Unsupervised Classification for Tiling Arrays: ChIP-chip and Transcriptome}
\end{center}

\begin{center}
Caroline B\'erard$^{1,2,*}$, Marie-Laure Martin-Magniette$^{1,2,3,4,5}$, V\'eronique Brunaud$^{3,4,5}$, S\'ebastien Aubourg$^{3,4,5}$ and St\'ephane Robin$^{1,2}$ \\
\end{center}

\noindent $^{1}$AgroParisTech, 16 rue Claude Bernard, 75231 Paris Cedex 05, France. \\
$^{2}$INRA UMR MIA 518, 16 rue Claude Bernard, 75231 Paris Cedex 05, France. \\
$^{3}$INRA UMR 1165, URGV, 2 rue Gaston Cr\'emieux, 91057, Evry Cedex, France. \\
$^{4}$UEVE, URGV, 2 rue Gaston Cr\'emieux, 91057, Evry Cedex, France. \\
$^{5}$CNRS ERL 8196, URGV, 2 rue Gaston Cr\'emieux, 91057, Evry Cedex, France. 
\begin{center}
$^{*}${caroline.berard@agroparistech.fr} \\ 
\end{center}



\vspace*{1cm}
\begin{abstract}
Tiling arrays make possible a large scale exploration of the genome thanks to probes which cover the whole genome with very high density until 2 000 000 probes. Biological questions usually addressed are either the expression difference between two conditions or the detection of transcribed regions. In this work we propose to consider simultaneously both questions as an unsupervised classification problem by modeling the joint distribution of the two conditions. In contrast to previous methods, we account for all available information on the probes as well as biological knowledge like annotation and spatial dependence between probes. Since probes are not biologically relevant units we propose a classification rule for non-connected regions covered by several probes. Applications to transcriptomic and ChIP-chip data of Arabidopsis thaliana obtained with a NimbleGen tiling array highlight the importance of a precise modeling and the region classification. \\
\end{abstract}

%
%

\begin{small}\textbf{Keywords}: Bivariate Gaussian mixture; Hidden Markov model; Tiling arrays; Unsupervised classification.\end{small}

\newpage

\section{Introduction}

\label{s:intro}

For 15 years, the study of large-scale genome is possible thanks to DNA microarrays. The probes, originally designed on genes, now cover the whole genome without \textit{a priori} knowledge of structural annotation: these are tiling arrays. The density is still increasing and companies now offer tiling arrays with 2 millions of probes. Thanks to technological advances and to the miniaturization of the support, tiling arrays have become a usual tool in biology laboratories. They make possible a large-scale exploration of the genome with a reasonable cost. Recently, the Next Generation Sequencing (NGS) technology revolutionize the domain because it directly products nucleotide sequences. However, like any new technology, it remains expensive and suffers for now from uncontrolled technical biases (\citealp{Oshlack}). It also raises new questions on read mapping or genome assembly. For all these reasons, tiling arrays, with a technology well controlled, remain widely used. They are a powerful tool to analyse all kinds of experiments and are used in a wide range of studies like DNA methylation, chromatin modification or transcription factor analysis with ChIP-chip experiments (\citealp{BuckLieb}), DNA copy number variation detection with CGH (\citealp{Pinkel}, \citealp{Snijders}) and survey of genomic transcriptional activities or transcript mapping with transcriptional experiments (\citealp{Mockler}, \citealp{Yamada}, \citealp{Hanada}). 




For comparative genomic hybridization, many different approaches exist for determining DNA copy number variations in CGH data like segmentation (\citealp{Hupe}, \citealp{Picard}) 
or Hidden Markov Models (\citealp{Fridlyand}, \citealp{Seifert-Colot}). 

For ChIP-chip experiments where the chromatin immunoprecipitation (ChIP) sample and the reference sample of genomic DNA are compared, the main goal is to detect regions enriched by ChIP. \citet{Johnson} proposed a Model-based Analysis of Tiling arrays (MAT) algorithm dedicated to Affymetrix arrays. MAT models the baseline probe behavior based on probe sequence characteristics and genome copy number. \citet{Li} proposed to model the behavior of each probe and then a 2-state HMM is used to estimate the enrichment probability at each probe location. In these two methods it is assumed that only a small proportion of probes is enriched by ChIP. This assumption is reasonable for ChIP-chip experiments dealing with transcription factor but not for histone modification or DNA methylation where a large enrichment is expected. \citet{Humburg} has suggested a parameter estimation procedure for robust HMM analysis of chromatin structure where several long regions of interest are expected. 
ChIP-chip data can also be seen as one signal along the genome when using the log-ratio between the intensities of the ChIP and the reference samples. Analyses are then usually done using a sliding window (\citealp{Cawley}) and statistic tests. \citet{Keles} and \citet{NTAP} have proposed respectively the Welch t-statistic and the non parametric Wilcoxon rank-sum method. The log-ratio is a relevant quantity only when its distribution is bimodal which is an ideal case not often obtained with real datasets. To overcome this problem \citet{ChIPmix} proposed modeling the distribution of the signal of the ChIP sample conditional to the reference signal by a bidimensionnal mixture of regression. \citet{ChIPmix} also proposed controlling a false positive risk. 
To perform comparative ChIP-chip study, two ChIP samples can also be directly compared with a bidimensionnal mixture model to study the differential enrichment (\citealp{Johannes}). The two samples then play symmetric roles.

Transcriptomic experiments may have two different purposes: the detection of transcribed regions or the study of gene expression across several conditions (also called differential analysis).
Most methods previously developed for tiling array transcriptomic data deal with the first purpose. Among them, some methods are based on probe-by-probe statistical tests (e.g. Fisher test developed by \citet{Halasz}) and others are based on segmentation methods such as \citet{Huber} or \citet{Zeller} or HMM (\citealp{Nicolas}). The incorporation of the annotation knowledge has also been proposed in a supervised framework (\citealp{Du}, \citealp{Munch}). 
Surprisingly few methods are devoted to the study of gene expression profiles across samples based on tiling arrays. Some methods aggregate probes within regions and then resume to hypothesis testing. The method gSAM (\citealp{gSAM}) is an extension of SAM, which models the differential expression of a given region by a constant piece-wise function. In the TileMap method (\citealp{TileMap}) each probe is used separately and a test statistic is proposed, based on a hierarchical empirical Bayes model. \\

\begin{figure}[!ht]
 \centering
 \includegraphics[scale=0.3]{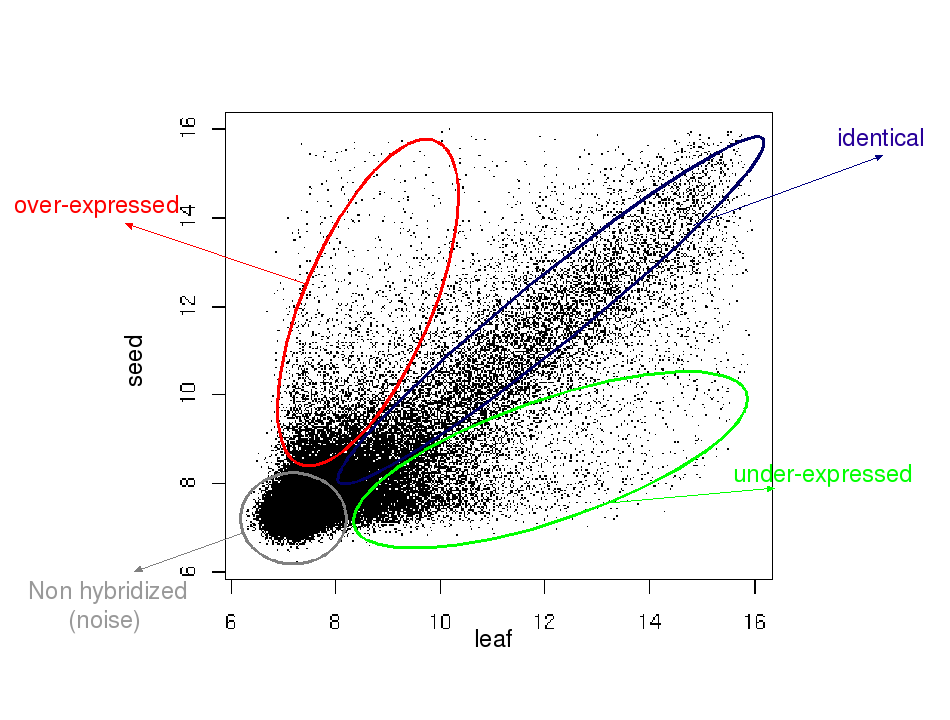}
 \caption{\label{4gpes}Schematic explanation of the 4 groups to consider when comparing two samples. Example of transcriptomic data.}
\end{figure}

In this article, we consider simultaneously the expression difference between two conditions and the detection of transcribed regions with an unsupervised classification point of view. We study the difference between two ChIP samples or between two transcriptomic samples. Comparing the two samples requires distinguishing four different biologically interpretable groups of probes: a group with similar behavior in both samples, a group with higher intensity in the first sample than in the second sample, a symmetric group with higher intensity in the second sample and a last group with low intensity in both samples which can be viewed as noise, corresponding to the non-transcribed regions (cf Figure \ref{4gpes}). 

\begin{figure}[!ht]
 \centering
 \includegraphics[scale=0.5]{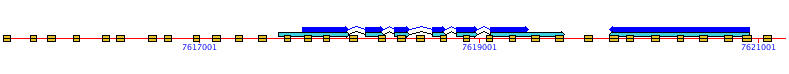}
 \caption{\label{annot}Example of genome annotation. Grey squares: probes; black arrows: known genes. The arrows correspond to exons and the fine lines between arrows correspond to introns.}
\end{figure}

A parametric classification method based on multivariate mixture models permits a direct comparison of two (or more) samples. Moreover this differs from a log-ratio based study which does not distinguish the group of identical behavior from the noise group. Therefore our method directly models the intensities of the two samples for all probes. In contrast to previous methods, we consider all the available information: the intensity of the two signals, the position of the probe along the genome and its structural annotation. The position of the probe is important because there is a signal dependence between adjacent probes due to the high resolution of tiling arrays. Structural annotation informs us about the location of the probes in intergenic, exonic or intronic regions (see Figure \ref{annot}, screen capture of FLAGdb++ (\citealp{flagdb})). This must be accounted for as, in a transcriptomic experiment, probes annotated as exonic are more likely to be hybridized whereas intergenic or intronic (non-coding) probes should be mainly in the noise group. We use a 4-state heterogenous hidden Markov model with bidimensional Gaussian emission densities to gather all this information. 
Finally since genome annotation is an on-going process with possible errors, we will discuss the relevance of its use for each specific application.\\

%

Most methods provide probe-by-probe results. As for the classification purpose, the HMM provides an answer for each probe, via the posterior probabilities. However probes are not relevant units from a biological point of view. Although HMM are widely used for classification problems in genomic data, the classification of biologically interesting regions is not a common practice. To get a result by region, the most common used method is a sliding window approach where the probe signals are merged \textit{a priori}. Another method proposed by \citet{Li} is to defined a region as at least two probes with positive log-odds enrichment value in ChIP sample and at least one probe with log-odds enrichment value lower than $-15$ in the control sample. But these methods do not deal with regions covered by several  non-adjacent probes, such as genes with exons and introns. We propose a new solution deriving a posterior probability for a region given \textit{a priori} with arbitrary structure (like a non-connected region) and also a procedure of gene classification which allows to get quickly a list of differentially expressed genes. This calculation by region really improves results than deducing a result from probe classifications \textit{a posteriori}.  \\


The article is organized as follows. The statistical model is described in Section \ref{model}. The inference is given in Section \ref{inference}. Section \ref{classif} describes the classification method for a probe and a gene. In Section \ref{appli}, we discuss the different sub-models of the method and apply them on NimbleGen tiling arrays for transcriptomic and ChIP-chip data of the plant \textit{Arabidopsis thaliana}. The main conclusions and some possible extensions are discussed in Section \ref{discu}. 
The method is implemented in R and C and is available upon request.

\section{Methods}
\label{method}
We propose a non-supervised classification model to compare the intensities of the two samples hybridized on the array for each probe. It accounts for all available information for each probe: the two intensities to be compared, the position of the probe along the chromosome and the current annotation of the probe (for example exonic, intronic, intergenic, transposable element, etc). As recommended in \citet{Johannes}, our method does not deal with the usual log-ratio but rather considers the two intensities separately and the joint signal of the two samples is modeled.

\subsection{Model}
\label{model}

For probe $t$, we denote
\begin{itemize}
\item  $X_{t}=(X_{t1},X_{t2})$ the log-intensities for both samples,
\item $C_t \in \{1, ... P\}$ the annotation category,
\item $Z_t \in \{1, ...K\}$ the unknown status. 
\end{itemize}
In our case, $K = 4$, Groups 1 and 2 will refer respectively to \textquoteleft noise' and \textquoteleft identical' probes, whereas Groups 3 and 4 will refer to differentially hybridized probes. 
To account for the dependence between adjacent probes, we assume that the process $\{Z_t\}$ is a first order Markov chain with heterogenous transition $\pi^p$ depending on the annotation category:
\begin{equation*}
 P(Z_{t}=l|Z_{t-1}=k, C_{t}=p) = \pi_{kl}^{p}
\end{equation*}
The proportion of observations in each group for each annotation category is given by the stationary distribution $m^p$ of the corresponding transition matrix $\pi^{p}$. We then assume that the $\{X_{t}\}$ are independent conditionally to the $\{Z_{t}\}$ with distribution
\begin{equation*}
X_{t}|Z_{t}=k \sim \mathcal{N}(\mu_{k},\Sigma_{k}).
\end{equation*}
The parameters $\mu_k$ and $\Sigma_k$ of the Gaussian do not depend on the annotation category. \\

If there is no spatial dependence, the $\{Z_{t}\}$ are independent and distributed according to a multinomial of parameter $\pi_{k}^{p}$ where $\pi_{k}^{p}$ corresponds to the proportion of probes from group $k$ in annotation category $p$. If there is no annotation and no spatial dependence, the model comes down to a mixture model with four components. All these sub-models are discussed in Section \ref{appli}.

\subsection{Inference}
\label{inference}

We use the parametrization proposed by \citet{Banfield} which enables us to characterize geometric properties of the Gaussian density (volume, shape, orientation). This parametrization considers the eigenvalue decomposition of the variance matrix of the group $k$: 
\begin{equation*}
\Sigma_{k} = D_{k} \Lambda_{k} D'_{k}.
\end{equation*}
The matrix $\Lambda_{k}$ describes both the volume and the shape of the ellipse associated with the Gaussian distribution. The matrix $D_{k}$ decribes the orientation of this ellipse. A similar decomposition of variance matrix is studied by \citet{Celeux} in the Gaussian mixture context and is implemented in the Mixmod software (\citealp{Mixmod}) and in the Mclust R package (\citealp{Mclust}). In their approach, each term of the decomposition is either equal in all groups or specific to each group. \\
In our case we need an intermediate modeling. By definition groups 1 and 2 should have the same orientation (see Figure \ref{4gpes}), which implies that $D_1 = D_2$. Furthermore the dispersion around the main axis is expected to be similar in all groups, which amounts to fixing the second eigenvalue of $\Sigma_{k}$ for all groups. This can be summarized as
\begin{eqnarray*}
 \Sigma_{k} & = & D_{k}\Lambda_{k}D'_{k}, \qquad  \mbox{for} \ k = 1,...,4; \\
 D_{1} &  = &  D_{2}; \\
 \Lambda_{k} & = & \begin{pmatrix}
                 u_{1k} & 0 \\
 		0 & u_{2}
                \end{pmatrix}, \qquad  \mbox{ with } u_{1k}>u_{2}, \mbox{for } k = 1,...,4.
 \end{eqnarray*}

The parameters $\{\pi^p\}$, $\{\mu_k\}$, $\{D_k\}$, $\{u_{1k}\}$ and $u_2$ are estimated using the EM algorithm. The E-step is achieved with the Forward-Backward algorithm (\citet{Baum}, \citet{Rabiner}). This model requires a specific M-step to satisfy the prescribed constraints on the variance matrices (see Appendix for formulas). These constraints cannot be satisfied with Mixmod or Mclust. In \citet{Johannes} the constraints are related to the means which assumes a strong symmetry in the distribution of the data. \\ 

\subsection{Posterior probabilities for a region}
\label{classif}

The posterior probability for each probe to belong to each group
$$
\tau_{tk, X}=P(Z_{t}=k|X), \qquad \text{where } X = \{X_t\},
$$
is obtained as a by-product of the Forward-Backward algorithm and can be used for probe classification.

As explained above, the probe may not be the relevant biological entity and we would rather look at the status of a region like a gene or a transposable element. We define a region as a set of probes, that can be decomposed into sub-regions of adjacent probes. As a reference to the gene structure and without loss of generality, we will refer to these sub-regions as \textquoteleft exons' and to the spaces between them as \textquoteleft introns'. In eukaryotic genes, exons correspond to coding regions that are spliced together in the transcript to become the mRNA, after removal of introns, which are not expressed. We define the posterior probability for such a region $g$ to belong to group $k$ as the probability for all its probes to belong to group $k$:
\begin{equation} \label{Qgk}
  Q_{gk, X} = P(\forall t \in g, Z_{t} = k | X,C) \\
\end{equation}
A region is covered by several probes and our definition considers the case of a homogeneous region, that is when all probes have the same status. \\
We compute this probability for a gene $g$ with $Q$ exons (and $Q-1$ introns). We denote $e_q$ the position of the first probe of exon $q$ and $i_q$ the position of the first probe of intron $q$; thus $i_q-1$ refers to the last probe of exon $q-1$. As convention, we denote $i_Q$ the position of the first probe after the end of the gene. We also denote $X_u^v = \{X_t\}_{u \leq t \leq v}$. We get
\begin{eqnarray*}
Q_{gk,X} & = & P(\forall t \in g, Z_{t}=k|X, C) \\
& = & P(Z_{e_{1}}=k|X_{1}^{e_{1}}) \times \left(\prod_{t=e_{1}+1}^{i_{1}-1} A_{k,t}\right) \times \prod_{q=2}^{Q-1}\left( B_{k,q} \times \prod_{t=e_{q}+1}^{i_{q}-1} A_{k,t}\right) \\
& & \times B_{k,Q} \times \left(\prod_{t=e_{Q}+1}^{i_{Q}-2} A_{k,t}\right) \times P(Z_{i_{Q}-1}=k|Z_{i_{Q}-2}=k, X_{i_{Q}-1}^{n}), \\
\text{where} \quad A_{k,s} & = & P(Z_{s}=k|Z_{s-1}=k,X_{s}, C), \\
B_{k,q} & = & P(Z_{e_{q}}=k| Z_{i_{q-1}-1}=k,X_{i_{q-1}}^{e_{q}}, C),
\end{eqnarray*}
where $C = \{C_t\}$. All these terms can be calculated with the Forward recursion of the Forward/Backward algorithm. 




Note that the sum of the $Q_{gk, X}$ for $k\in\{1, ...4\}$ is not equal to one, as all probes from a same gene may not have the same status. Changing the list of exons associated to a gene allows us to account for alternative splicing or exclude the last exon for which the expression level could be lower due to the labeling protocol (\citealp{Nicolas}). 

\section{Applications}
\label{appli}

We now illustrate the use of the proposed modeling on both ChIP-chip and transcriptomic data. All experiments have been carried out on a two-color NimbleGen array of about 700~000 probes designed to insure a constant hybridization temperature. For each dataset two biological replicates are available, for which hybridizations are performed in dye-swap. The normalization step is done by averaging on the dye-swap the two signals of each technical replicate to remove the gene-specific dye bias (\citealp{Mary-Huard}). Analyses are performed per chromosome on the normalized data. 

\subsection{ChIP-chip dataset}
\label{ChIPchip}

We analyse the data from a histone modification (H3K9me2) study in \textit{Arabidopsis thaliana} for a wildtype and a mutant (polIV). We directly compare the ChIP samples of the wildtype and the mutant to study their difference in methylation.
The methylation mainly affects transposable elements but also large adjacent regions (\citealp{Humburg}). Therefore we expect to find enriched probes both in the transposable elements and in wide neighboring regions. As the methylation does not affect a specific annotation category, the standard annotation information is not useful to detect enriched probes. This suggests using a HMM model without the annotation knowledge. \\
The histone methylation under study is known to be weakly present in the genome and the mutant is known to have a loss of methylation compared to the wildtype (\citealp{Bernatavichute}). We find consistent results as shown by the estimated proportions in each group given by our model: 43\% noise, 21\% identical, 22\% lost in mutant, 14\% gain in mutant. \\
The studied histone modification is also known as a heterochromatin mark. Most regions covered by H3K9me2 are adjacent and cover several megabases in pericentromeric regions or in interstitial heterochromatin regions (tightly packed form of chromatin) as the knob of chromosome 4, but there are also smaller regions (islands of heterochromatin) located in euchromatin (lightly packed form of chromatin) and covering mainly transposable elements (\citealp{Bernatavichute}). The results obtained using our method corroborate this information: 91.3\% of probes in heterochromatin are methylated whereas only 49.5\% of probes in euchromatin are methylated. In heterochromatin, 82\% of probes have identical behaviour between wildtype and mutant whereas only 9.5\% of probes are identical in euchromatin. Moreover 56\% of methylated probes cover transposable elements or a 500 base-pair (bp) surrounding region. 
The transition probabilities provide insights about the length of regions from each group through mean sojourn time. The average size of the binding sites is 14.3 probes (corresponding to 3289\textit{bps}) for the identical group, 4.5 probes (1035\textit{bps}) for the group with lost in mutant, 3.7 probes (851\textit{bps}) for the group enriched in mutant and 7.7 probes (1771\textit{bps}) for the noise group. \\
These calculations show that impoverished or enriched regions are three times smaller than regions with identical behaviour between wildtype and mutant. Moreover, the transposable elements are 2 to 3 times smaller in the euchromatin compared to the heterochromatin. This suggests that most of the methylation losses of the mutant occur in transposable elements from the euchromatin.\\
The transposable element META1 (located between positions 5326458 and 5331580 on the chromosome 4) is known to have a loss of methylation in the mutant. The regulatory region of META1 is located at the beginning of the transposable element with small RNAs which are involved in methylation process. Our method declared the first half of the probes covering META1 (near the start position) in the group where methylation is lost. The other probes are declared identically methyled between the two samples. This example shows the advantage of the high resolution of the tiling array. \\

\paragraph{Comparison with the models of \citet{Johannes}}
\label{ComparisonJohannes}

\begin{figure}[!ht]
 \centering
 \includegraphics[width=13cm,height=9cm]{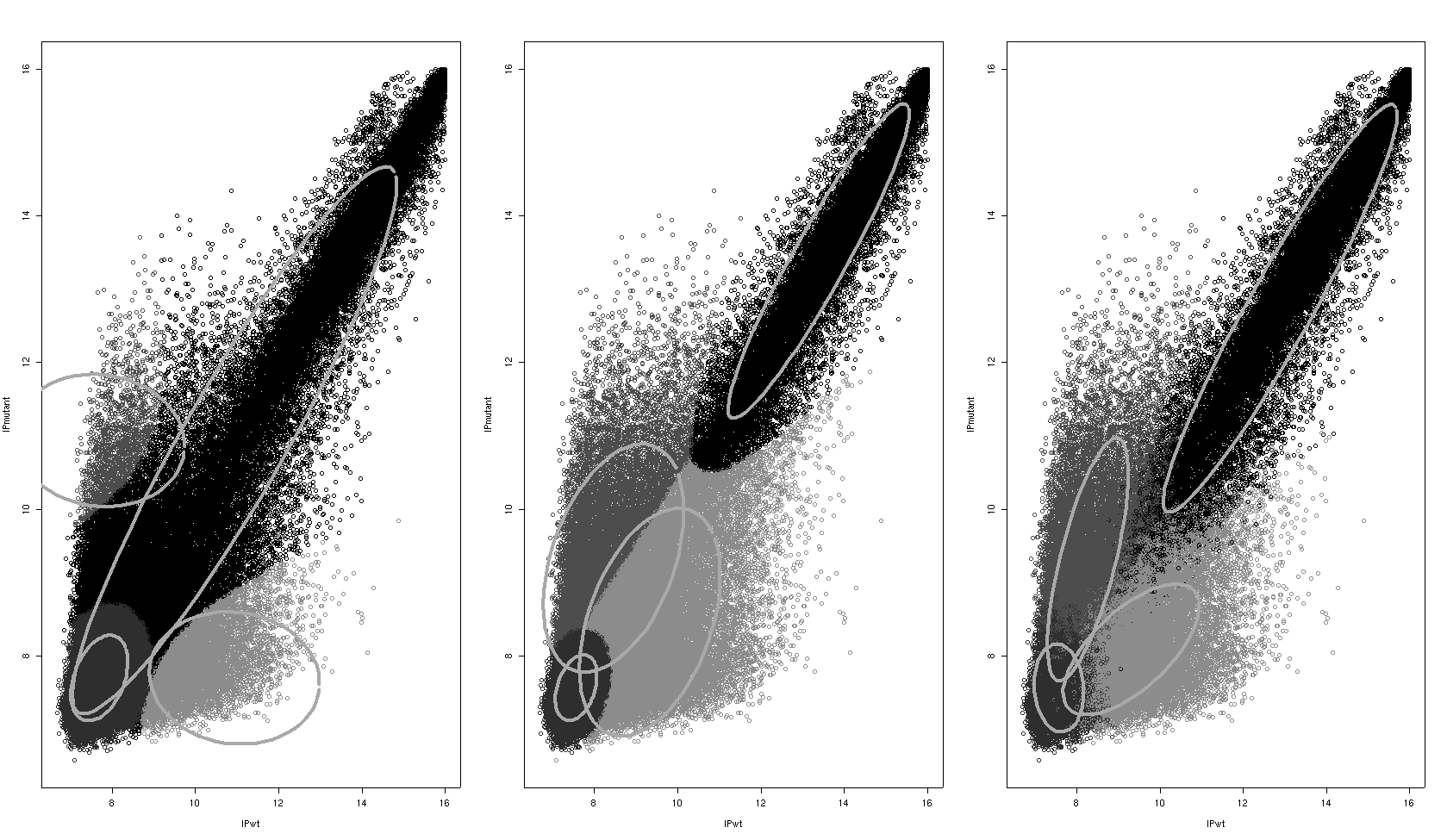}
 \caption{\label{model_Johannes}Classification comparison between our HMM model and the models 2 and 3 of \citet{Johannes} on H3K9me2 dataset. Left: Model 2, Middle: Model 3, Right: our HMM model.}
\end{figure}

As in Section \ref{model}, Groups 1 and 2 refer respectively to \textquoteleft noise' and \textquoteleft identical' probes, whereas Groups 3 and 4 correspond to differentially enriched probes. 
\citet{Johannes} proposed two models. A full-switching model (Model 2) where the component means are constrained as follows: $\vec \mu_{1}= (\mu_{1}, \mu_{1})$, $\vec \mu_{2}= (\mu_{2}, \mu_{2})$, $\vec \mu_{3}= (\mu_{2}, \mu_{1})$, $\vec \mu_{4}= (\mu_{1}, \mu_{2})$ and the covariances matrices of Groups 3 and 4 are equal ($\Sigma_{3}=\Sigma_{4}$). Model 3 is a flexible-switching model similar to the
full-switching model (Model 2) with less constraints: $\vec \mu_{3}= (\mu_{4}, \mu_{3})$ and $\vec \mu_{4}= (\mu_{3}, \mu_{4})$. 
We compared our model with their models on the H3K9me2 dataset. Model 2 leads to a smaller proportion of differentially enriched regions (7.8\% lost in mutant and 1.2\% gain in mutant, see Figure \ref{model_Johannes}) than our model (respectively 22\% and 14\%). The transposable element META1 that is declared differentially enriched with our model (see above) is found in the identical group according to their Model 2. The classification of Model 3 seems not to be suitable for probes with similar intensities between 8 and 10 where more probes are expected to be declared in the identical group (see Figure \ref{model_Johannes}). In conclusion, it seems that the independence assumption, the symmetrical constraints on the means and, the equal variances for the differentially enriched probes lead to a too simple model for analyzing such data. These two models do not also fit well the transcriptional dataset defined in Section \ref{transcriptome} (results not shown). 

\subsection{Transcriptional dataset}
\label{transcriptome}

We now study the gene differential expression between the leaf and the seed 10 days after pollination of the plant \textit{Arabidopsis thaliana}. First we compare the 4 sub-models, second we present the results by gene, then we consider the detection of new transcribed regions. In the results, over-expressed (under-expressed) refers to probes with higher (smaller) signal in the seed.

\subsubsection{Comparison of the 4 models}
\label{results}
The model presented in Section \ref{model} uses all available information and is referred to as model $\mathcal{M}_{4}$. For the annotation, $P = 3$ categories are considered: intergenic, intron and exon, and only exons are expected to be transcribed. Model $\mathcal{M}_{4}$ can be simplified if either the structural annotation (Model $\mathcal{M}_{2}$) or spatial dependence between probes (Model $\mathcal{M}_{3}$) is not taken into account. The model $\mathcal{M}_{1}$ is the simplest model with neither annotation nor spatial dependence. It comes down to an independent mixture model. The constraints on the variance matrices detailed in Section \ref{inference} are kept in all models. 
Table \ref{vraisemblance} presents the fit of the four models for chromosome 4. The full model $\mathcal{M}_{4}$ achieves the best BIC criterion, suggesting that all available information should be taken into account. ICL (\citealp{ICL}) is an alternative model selection dedicated to classification purposes. According to ICL, the model $\mathcal{M}_3$ (without HMM) should be chosen, meaning that the annotation information somehow contains the information about the spatial dependence. Nevertheless, since the difference between $\mathcal{M}_{3}$ and $\mathcal{M}_{4}$ is small in terms of ICL and the number of parameters is far smaller than the number of observations, we use the model $\mathcal{M}_{4}$ to compare the two transcriptomic samples. 


\begin{table}
\caption{\label{vraisemblance}Fit of the 4 models.}
 \centering
\begin{tabular}{lccccc}
\hline
 & $\mathcal{M}_{1}$ & $\mathcal{M}_{2}$ & $\mathcal{M}_{3}$ & $\mathcal{M}_{4}$ \\
\hline
number of parameters & 18 & 30 & 24 & 60 \\
$-2$ log-likelihood & 406248 & 371308 & 373282 &  356616 \\
BIC & 406457 & 371656 & 373561 & 357312 \\
ICL & 436185 & 416529 & 399975 & 400330 \\ 
\hline
\end{tabular}
\end{table}


As expected, intergenic probes mostly belong to the noise group (84\%) and few belong to expressed groups: 9\% in the under-expressed group and 6\% in the over-expressed. Intronic probes display a similar, although different, repartition: 60\% noise, 7\% identical, 24\% under-expressed and 9\% over-expressed (cf Section \ref{detection} for discussion about expressed probes in intergenic and intronic regions).  As expected, most exonic probes (78\%) belong to the expressed groups: 41\% identical, 23\% under-expressed and 14\% over-expressed. The transition matrices for the intronic and intergenic categories are very similar (not shown): whatever the status of probe $t$, probe $t+1$ has between 70\% and 95\% chance of being noise. This is different for the exonic probes where the transition matrix has high probabilities on the diagonal meaning that probe $t+1$ has high probability (80\% to 90\%) of having the same status as probe $t$. All these results seem to be coherent with what is expected for transcriptomic data.

\subsubsection{Gene classification}
 \label{gene}

We now consider the classification of each gene. To this end, we compute the posterior probability $Q_{gk, X}$ defined in Equation \eqref{Qgk}. We propose to classify the genes via a two-step procedure. The probability for a gene to be homogeneous whatever the status is $\sum_{k}Q_{gk,X}$. We first verify if the gene has homogeneous status by considering a ratio similar to a Bayes factor: $\sum_{k}Q_{gk,X}/\sum_{k}Q_{gk}$, where $Q_{gk} = P(\forall t \in g, Z_t = k | C)$ is the non-conditional version of $Q_{gk, X}$, which is for a gene
$$
Q_{gk} = m^E_k \times (\pi^E_{kk})^{\sum_{q=1}^Q (i_q-e_q)-1} \times \prod_{q=1}^{Q-1} [(\pi^I)^{e_{q+1}-i_q}]_{kk} \ , \\
$$
where superscripts $E$ and $I$ refer to the exonic and intronic categories, respectively.  \\
As its computation involves a product of probabilities with as many terms as the number of exonic probes in the gene, $Q_{gk, X}$ goes to zero for long genes. The ratio with $Q_{gk}$ does not correct this effect, therefore we apply an additional linear correction on the log-ratio with respect to the length of exons and the number of exons in the gene. \\
We define this corrected log-ratio as \textit{unistatus} value which is a tool for decision support. \\
If the homogeneous assumption seems verified, the second step is to calculate the conditional posterior probability $Q_{gk,X}/\sum_{l}Q_{gl,X}$ to assign the gene to the group $k$ for which this posterior probability is the highest. \\
We found 81\% of genes which have an unistatus value higer than 0 (corresponding to 1736 genes). Among these 1736 genes, 920 are declared identically expressed in the seed and in the leaf, 318 are declared under-expressed in the seed and 181 are declared over-expressed in the seed. We focus our analysis on 8 genes clearly identified as preferentially transcribed in seeds using graphical outputs of Genevestigator which is a database of transcriptome analysis results (\citealp{genevestigator}). Among the 8 genes, 7 have an unistatus value higer than 0 and are declared over-expressed in seed with our calculation.\\

\subsubsection{Detection of new transcripts}
\label{detection}

\begin{figure}[!ht]
 \centering
 \includegraphics[scale=0.4]{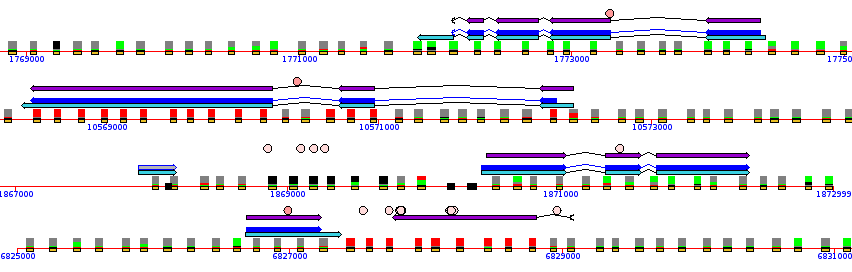}
 \caption{\label{fig:03}Presentation of results in Flagdb++. Circles represent EST, white arrows are official TAIR annotation genes and black arrows represent Eugene genes. The small rectangles are probes colored according to their status: light if not expressed, dark if expressed. On the first 2 lines, there are 2 expressed genes covered by a majority of probes with the same status except the intronic probes which are reasonably declared as not expressed. The last 2 lines present expressed probes where there are no official genes but the signal coincides with EST and/or Eugene genes.}
\end{figure}

Although our model is built for the comparison of two samples, it also allows the detection of previously unknown transcription sites thanks to the high resolution of the tiling array. To this aim, the model without annotation seems more suitable, since we are bringing it into question.
A lot of regions with expressed probes are found in intergenic regions: 1328 small regions with 2 or 3 consecutive expressed probes, 185 regions with 4 or 5 consecutive expressed probes and 90 regions with more than 5 consecutive probes (including 25 regions with more than 10 consecutive probes).
For the 90 regions with more than 5 consecutive probes, we check with other annotation information like Expressed Sequence TAG (EST) or genes predicted by the Eugene software (\citealp{Eugene}) which are not yet in the official TAIR annotation. We found 39 regions matching with annotation like small RNA, rRNA, tRNA, including 12 regions corresponding to a coding sequence defined in Eugene and 10 corresponding to transcriptional units recently annotated due to the presence of EST.
The Figure \ref{fig:03} (from FLAGdb++ (\citealp{flagdb})) shows examples of results for two annotated genes and also for two expressed regions which correspond to EST and Eugene genes. 
Moreover the obtained results show many other interesting things, such as surprisingly many transcriptions in the introns in 5'UTR (40\% of intronic probes declared expressed in Section \ref{results}). This seems to be consistent with a recent article of \citet{Cenik} assuming a functional role of 5'UTR short introns. \\


\section{Discussion}
\label{discu}

Tiling array is a powerful technology which requires adapted statistical methods to deal with the large quantity and the variability of the data.
We focus on the comparison of two samples from transcriptomic or ChIP-chip experiments and also on the detection of transcribed regions by directly modeling the joint distribution of the two sample intensities. We consider all the available information from the probes: the intensity of the two signals, the dependence between neighboring probes and the structural annotation. The annotation knowledge is very useful information with the aim of classification because of the intrinsic difference between exonic or intergenic probes. This method can be used for ChIP-chip or transcriptomic data whenever there are two conditions to compare. Both one-color and two-color tiling arrays can be analysed. Applications on \textit{Arabidopsis thaliana} tiling array show the ability of the model to interpret the data and provide a new insight of gene expression or gene expression control as well as new biological hypothesis. \\
\textit{Arabidopsis thaliana} is a model plant with a very well-known genome annotation but the annotation is not available or unreliable for many organisms. That is why the sub-models are also useful. 
The model without annotation allows us not to be limited by the quality of the available annotation and this model is also useful to detect new genes to improve the current official annotation. \\
This work raises also the question of classification. The results are given by probe and by region. We compute a posterior probability by region and we propose a procedure of region classification. The most common regions are the genes which are non-connected regions, but any other region can be defined. It would be interesting to control the False Discovery Rate, \textit{i.e.} the expected proportion of misclassifications, in the case of having 4 groups and under dependence hypothesis, and also for the results given by region. 
\\

%
%

%
\section*{Acknowledgements}
The authors thank F. Roudier and V. Colot from IBENS for providing ChIP-chip data and for helpful discussions to biological interpretation, and S. Derozier and A. Lecharny from URGV for their help in bioinformatic analysis. The authors are grateful to S. Balzergue from URGV for providing transcriptomic tiling array data. This work was funded by MIA, GAP and MICA departments of INRA.



\appendix

\section{Appendix: Computation of the estimates of $D$ and $\Lambda$.}

\label{Appendix}

Recall $X_{t}=(X_{t1},X_{t2})$ the log-intensities for both samples, t varies from 1 to n, where n is the total number of observations.
Let $\bar X_{k} = \frac{\sum_{t=1}^{n} \tau_{tk,X}X_{t}}{n_{k}}$, where $n_{k}=\displaystyle{\sum_{t=1}^{n}} \tau_{tk,X}$. \\
Let $ W_{k} = \displaystyle{\sum_{t=1}^{n}}\tau_{tk, X}(X_{t}-\bar X_{k})(X_{t}-\bar X_{k})'$ be a matrix like $\begin{pmatrix}
w_{1k} & w_{2k} \\
w_{2k} & w_{4k}
\end{pmatrix}$ and
$\Lambda_{k}  = \begin{pmatrix}
u_{1k} & 0 \\
0 & u_{2}
\end{pmatrix},\\ \mbox{ with } u_{1k}>u_{2}, \mbox{for } k = 1,...,4.$

The maximum likelihood estimator of the orientation matrix \textit{D} identical for the first two components with the same orientation is in the form of $\begin{pmatrix}
\hat d & -\sqrt{1-\hat d^{2}} \\
\sqrt{1-\hat d^{2}} &   \hat d 
\end{pmatrix}$,
where $\hat{d}$ is the minimum of the function:
\begin{equation*}
 f(d) = \sum_{k=1}^{2}\left\lbrace \frac{d^{2}w_{1k}+2w_{2k}d\sqrt{1-d^{2}}+w_{4k}(1-d^{2})}{u_{1k}} + \frac{d^{2}w_{4k}+2w_{2k}d\sqrt{1-d^{2}}+w_{1k}(1-d^{2})}{u_{2}} \right\rbrace 
\end{equation*}
The estimator of $\hat{d}$ is defined by:
\begin{equation*}
\widehat{d^{2}} - \frac{1}{2} \ = \ \pm \frac{N_{1,4}}{2\left[ \left\lbrace N_{1,4} \right\rbrace ^{2} + 4\left\lbrace N_{2} \right\rbrace ^{2}\right]^{1/2}} , \  \  \mbox{with }  \hat{d} > 0, \\ 
\label{D}
\end{equation*}
where $N_{1,4} = \displaystyle{\sum_{k=1}^{2}}(w_{1k}-w_{4k})(u_{2}-u_{1k})/u_{1k}u_{2}\ $ and $\ N_{2} = \displaystyle{\sum_{k=1}^{2}}(w_{2k})(u_{2}-u_{1k})/u_{1k}u_{2}$.


Let $B_{k}$ be a matrix defined by $B_{k} = D'_{k}W_{k}D_{k}$ like $\begin{pmatrix}
b_{1k} & b_{3k} \\
b_{4k} &  b_{2k} 
\end{pmatrix}$. 
The maximum likelihood estimator of $\Lambda_{k}$ is in the form of $\begin{pmatrix}
\hat u_{1k} & 0 \\
0 &  \hat u_{2} 
\end{pmatrix}$, where 
\begin{equation*}
 \left\{
\begin{array}{rl}
\hat u_{1k} & = b_{1k}/n_{k} \\
\hat u_{2} & = \sum_{k=1}^{4}b_{2k}/n.
\end{array}
\right. \\
\label{Lambda}
\end{equation*}


\end{document}